\documentclass{article}[11pt]
\usepackage[utf8]{inputenc}
\usepackage[paper=a4paper,dvips,top=3cm,left=2.4cm,right=2.4cm,foot=1cm,bottom=3cm]{geometry}
\usepackage{hyperref}
\usepackage{amssymb}
\usepackage{amsmath}
\usepackage{authblk}
\usepackage{amsthm}
\usepackage{diagbox}
\usepackage{graphicx}
\usepackage{epstopdf}
\usepackage{url}
\usepackage{color}
\usepackage{float}
\usepackage{booktabs}
\usepackage{algorithm}
\usepackage{algorithmic}
\usepackage{url}
\usepackage{mathrsfs}
\usepackage{amsopn}
\usepackage{amssymb}
\usepackage[section]{placeins}
\usepackage{subcaption}

\begin{document}

\title{\huge Flow-Level Packet Loss Detection via Sketch Decomposition and Matrix Optimization}

\author[1]{Zhenyu Ming}
\author[2]{Wei Zhang}
\author[1]{Yanwei Xu}
\affil[1]{Theory Lab, Central Research Institute, 2012 Labs, Huawei Technologies Co., Ltd., Hong Kong}
\affil[2]{Department of Mathematical Sciences, Tsinghua University, Beijing 100084, China}

\date{}
\maketitle

\begin{abstract} 
For cloud service providers, fine-grained packet loss detection across data centers is crucial in improving their service level and increasing business income. However, the inability to obtain sufficient measurements makes it difficult owing to the fundamental limit that the wide-area network links responsible for communication are not under their management. Moreover, millisecond-level delay jitter and clock synchronization errors in the WAN disable many tools that perform well in data center networks on this issue. Therefore, there is an urgent need to develop a new tool or method. In this work,  we propose SketchDecomp, a novel loss detection method, from a mathematical perspective that has never been considered before. Its key is to decompose sketches upstream and downstream into several sub-sketches and builds a low-rank matrix optimization model to solve them.  Extensive experiments on the test bed demonstrate its superiority.

{\bf keywords: packet loss; sketch decompostion; matrix optimization} 

\end{abstract}


\section{Introduction}
Packet loss is common in various networks, and its rate is an important metric that characterizes network performance \cite{de2008one}. For cloud service providers(CSPs) with multiple data centers, sufficient knowledge of packet loss, especially in the forward path between data centers, contributes to improving the quality of service and avoiding violating service-level agreements with customers. Hence, reliable, fine-grained packet loss detection, that is, knowing the packet loss of each flow in each short time window, is increasingly valued and has been desired by them for a long time. However, restricted by the non-visibility of communication links, it is usually a challenging task. Because of the long geographical distance, the data centers are typically connected by a backbone network managed by Internet service providers. It means communication on network links is not under the control of CSPs. They have only access to monitor traffic on egress and ingress devices in their own data centers, not at any point on the wide area network link. The incomplete measurements fundamentally limit the generation and development of solutions.

Many popular monitoring tools proposed before are not up to the task, at least not as fine-grained. EverFlow \cite{zhu2015packet} and s-Flow \cite{wang2004sflow} mirror packets as they pass through the switching device to record information and then compare those mirrored upstream and downstream to identify packet loss. The high bandwidth and storage overhead force them to sample only some packages for mirroring and thus fail to meet the accuracy requirements. NetFlow \cite{claise2004cisco} maintains a counter for each flow on the switching device to realize flow-level measurements. FlowRadar \cite{li2016flowradar}, on top of NetFlow, introduces encoded flow sets to save the memory overhead of switches and sets up a remote collector to decode the flows and counters network-wide. They both ignore the misalignment of upstream and downstream time windows caused by clock offset and variable transmission delay, leading to biased packet loss analysis in a short time window. LossRadar \cite{li2016lossradar} is a seminal work, which marks the time window ID in the packet header upstream to identify the source of packets and thus counter the impact of time window misalignment. Despite its success in data center networks, it is not applicable and uneconomic to the backbone network we focus on. Specifically, the data center's egress devices and ingress devices are often non-programmable commercial switches, hard to add packet headers efficiently. Moreover, routers beyond the control of CSPs may not support forwarding non-standard packets with headers added. 

This paper proposes SketchDecomp, a novel detection method for packet loss between data centers. Based on sketch decomposition and matrix optimization, it does a high-accuracy job at the flow level without modification of packets or other hardware operations. Thus, it is easy to deploy and economical compared to previous methods. Concretely, in SketchDecomp, several count-min sketches \cite{cormode2005improved},  a particular data structure represented as a matrix, are deployed upstream and downstream to count flows in the send and receive windows. Given that packets sent may be branched to multiple receive windows due to delay jitter, we regard the sketches as the sum of several unknown sub-sketches. After some equality and inequality constraint are derived, a matrix optimization model following low-rank property is subsequently developed to solve these sub-sketches. In particular, we design a symmetric Gauss–Seidel alternating direction method in order to solve it efficiently. Finally, SketchDecomp identifies the packet loss and computes the loss rate by comparing the sub-sketches recovered and sketches observed. 

The rest of the paper is organized as follows: In Section 2, we give the precise problem statement and detail the deployment and optimization problem modeling of SketchDecomp. In Section 3. we present the sGS-ADMM algorithm to solve the optimization model and cover its convergence. Extensive experiments in Section 4
highlight the efficiency and accuracy of SketchDecomp. Finally, we conclude the work in Section 5.

\section{SketchDecomp}
In this section, we briefly describe the packet loss detection problem to be solved and then develop our proposed method.
\subsection{Problem Statement}
Consider communication between two data centers $DC_1$ and $DC_2$, bridged by a WAN link $L$.  Their clocks have been synchronized using certain synchronization methods of the WAN. In a certain period, streaming data consisting of several types of flows are continuously sent from upstream $DC_1$ to downstream $DC_2$ through $L$. The forward delay varies with time, and its rough range is known. Dividing the upstream time into multiple windows of length $T$, we aim to check for each flow whether the packet loss occurs in each window and estimate the packet loss rate accurately.

\subsection{Count-min Sketch}
\begin{figure}
    \centering
   \includegraphics[width=0.5\linewidth]{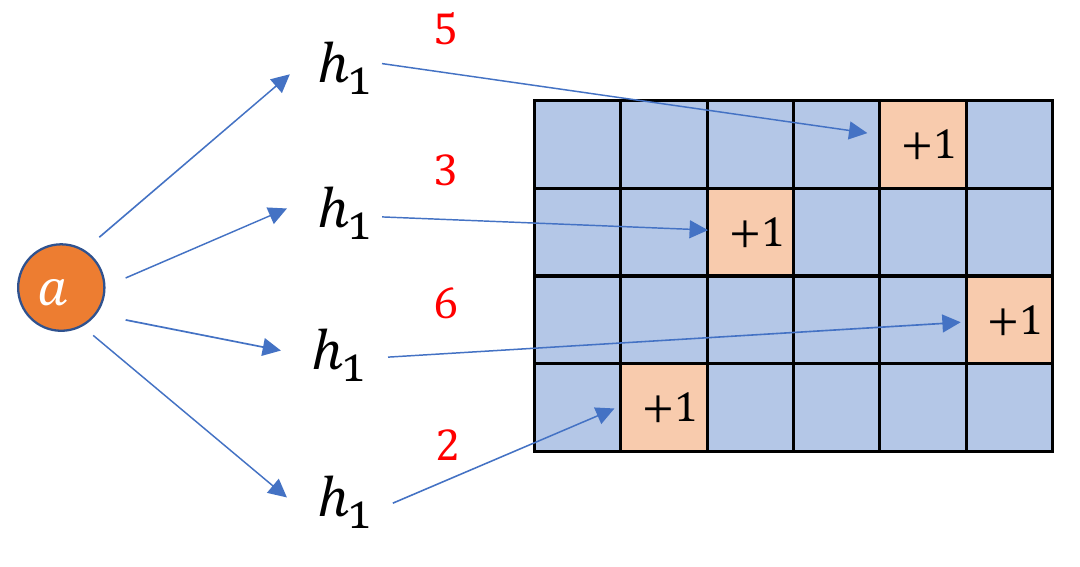}
    \caption{Sketch}
    \label{fig:sketch}
\end{figure}

The count-min sketch (CM sketch) is a particular data structure to count and record the frequency of each type of flow in stream data, with a certain confidence level. Unlike traditional hash tables, it utilizes multiple, say $d$, mutually independent hash functions $\{h_i\}_{i=1}^d$ to against the overcounting caused by collisions. Basically, CM-Sketch is a $d \times w$ matrix denoted as $C$, whose $i$-th row is the count created by $h_i$. There is an important property of $C$ that its row sums are all equal, i.e,
\begin{equation*}
    AC \mathbf{1} =\mathbf{0},
\end{equation*}
where $A$ reads 
\begin{equation*}
    A = \begin{bmatrix}
1 & -1 & & & & \\
 & 1 & -1 & &  &\\
 &  & \ddots & \ddots &\\
 & &  & 1 & -1 &
    \end{bmatrix}_{(d-1) \times d}
\end{equation*}
In the beginning, matrix $C$ is initialized to all zero. For each incoming element $a$, we hash it in turn by $\{h_i\}_{i=1}^d$ to get $d$ column indexes and then increase the value of the matrix $C(i,h_i(a))$ at the respective index by one, as shown in Figure \ref{fig:sketch}. After all flow elements have been viewed, The matrix will keep fixed and enable querying. The rule for querying is concise: The final estimated count of an element $a$ is the least value of the counts given by all hash functions, i.e., $\min_i h_i(a)$. 

\subsection{Sketch Deployment}
Same as upstream, we divide the downstream time into time windows of length $T$. Their start time is set to be later than upstream by the estimated minimum delay, because of the transmission delay.
     Then we deploy a CM sketch for each of $n$ time windows upstream and downstream for the preliminary information collection,  These $2n$ sketches have the same width $w$ and depth $d$ and adopt the same hash functions set. For the sake of notation, we denote the upstream sketch over the $k$-th time windows as $S_k$ and the corresponding downstream one as $R_k$.

    Generally, it is biased to conclude a packet loss in $k$-th windows by observing the difference $S_k$ and $R_k$ as delay jitter frequently occurs in WAN. Since we only consider the minimum transmission delay when setting downstream windows, the packet sent in the $k$-th windows that experiences a larger delay will be received in the $k+1$, $k+2$, or the later windows.  In fact, no matter how we set the downstream windows, we can't analyze packet loss in the way above.
\subsection{Sketch Decomposition}
Assume that the jitter range of the link delay does not exceed $m-1 \ (m<n)$ times the length of window $T$, then packets sent in the upstream $k$-th windows, if not lost, can only be branched into the downstream $k$-th, $(k+1)$-th,$\cdots$, $(k+m-1)$-th windows. Each branch as well as the dropped packets can construct a corresponding sub-sketch.  Let $M_{ki}$ be the sub-sketch constructed from the packets sent in the $k$-th window and received in the $(k+i-1)$-th window. $\phi_i$ is the sub-sketch representing the dropped packets of the $i$-th window. The upstream sketches have the following sketch decomposition:
\begin{equation*}
    S_k = \sum_{i=1}^{m} M_{ki}+\phi_k,\quad k=1, \cdots, n.
\end{equation*}
Without regard to $\phi_i$, it reads that
\begin{equation}
    S_k \geq \sum_{i=1}^{m} M_{ki},,\quad k=1, \cdots, n.
\end{equation}
For the downstream sketches, a similar decomposition holds:
\begin{equation*}
    R_k=\sum_{i=1}^m M_{k-i+1,i},\quad  k= m, \cdots,n.
\end{equation*}
Our concern is to solve all the $M_{ki}$ for $k=1,\cdots n, \ i=1,\cdots m$ in order to analyze packet loss by the discrepancy between $\sum_{i=0}^m M_{k,i}$, $S_k$. However, it is currently unsolvable for the reason that there are $n*m$ variable matrices but only $n-m$ equality constraints. 
\subsection{Low-rank Matrix Optimization}
In this part, we regard sketch decomposition as a matrix optimization problem and explore its low-rank property.

As we said before, in a time window, several types of flows are transmitted to the link $L$.
Usually, for a link of WAN, there is not a dramatic change in congestion for a short time.
 Therefore, we have enough to think that these flows have a similar delay distribution. In other words, each flow has a close proportion of packets branched into the same time window downstream. As a direct result, given $k \in \{1,\cdots,n\}$, the sub-sketches $M_{k1},M_{k2},\cdots M_{km}$ are proportional to each other. More formally, if we concatenate them into a large matrix $M$ as
 \begin{equation*}
     M = \left[M_{11}; M_{12};\cdots; M_{1m}; \cdots \cdots;M_{n1} ; M_{n2};\cdots ;M_{nm}\right],
 \end{equation*}
 $M $ will be a low-rank matrix of dimension $\lambda \times w$ where $\lambda = nmd$. 

The above considerations suggest that it is advisable to take minimizing the rank of $M$ as the optimization objective. So we give the following optimization model:
\begin{equation*}
    \begin{gathered}
         \min_M \quad \lVert M\rVert_*\\
    \text{s.t.} \quad  \left\{
    \begin{aligned}
         &\sum_{i=1}^{m} M_{k-i+1,i}=R_k,\quad  k= m, \cdots,n, \\
         &\sum_{i=1}^{m} M_{k,i} \leq S_k,,\quad k=1, \cdots, n,\\
         & AM_{k,i}\mathbf{1} = \mathbf{0}, \quad k= 1,\cdots,n, \quad i = 1,\cdots,m, \\
         &M \geq 0.
    \end{aligned}
    \right.
    \end{gathered}
\end{equation*}
where $\lVert\,\rVert_*$ refers to the nuclear norm. 

To be more generalized in form, we consider to write the constraint explicitly as a linear constraint on $M$. For this, we introduce three matrice $B_1 \in\mathbb{R}^{(n-m+1)d \times \lambda}, B_2 \in \mathbb{R}^{nd \times \lambda}, B_3 \in \mathbb{R}^{nmd \times \lambda}$ for the left-hand side, which satisfies
\begin{equation*}
    B_1 M = \begin{bmatrix}
        \sum_{i=1}^{m} M_{m-i+1,i}\\
        \cdots \\
        \sum_{i=1}^{m} M_{n-i+1,i}
    \end{bmatrix},\quad 
    B_2M = \begin{bmatrix}
        \sum_{i=1}^{m} M_{1i}\\
        \cdots \\
        \sum_{i=1}^{m} M_{ni}
    \end{bmatrix},\quad 
    B_3M = \begin{bmatrix}
    AM_{11}\\
    \cdots \\
    AM_{1m}\\
    \cdots\\
   AM_{n1}\\
    \cdots \\
    AM_{nm}\\
\end{bmatrix},
\end{equation*}
and two matrice $R \in \mathbb{R}^{(n-m+1)d \times \lambda} ,S \in \mathbb{R}^{nd \times \lambda} $ satisfying
\begin{equation*}
    R = \begin{bmatrix}
    R_{m+1} \\
    \cdots \\
    R_n
\end{bmatrix}, \quad
S = \begin{bmatrix}
    S_{1} \\
    \cdots \\
    S_{n}
\end{bmatrix}.
\end{equation*} for the right-hand side.
Then the model can be reformulated as 
\begin{equation}\label{eq:reformulated model}
         \min_M \ \lVert M\rVert_* \qquad
    \text{s.t.} \quad 
         B_1M = R, \ \  B_2M \leq S,\ \   B_3M \mathbf{1}=\mathbf{0},\ \  M \geq 0.
\end{equation}

\section{sGS-ADMM Algorithm}
In recent years, many studies \cite{han2017,chen2019,chang2020,he2018,li2019,CHEN2017} recommend
alternating direction method of multipliers algorithms for solving model \eqref{eq:reformulated model}. Compared to the classic interior-point methods \cite{QPIP}, these ADMM-type algorithms require fewer computational resources, e.g., memory and time. In what follows, we propose a symmetric Gauss-Seidel ADMM (sGS-ADMM) algorithm to solve the model.

The Lagrangian function of model \eqref{eq:reformulated model} is 
\begin{equation*}
    L(M) = \lVert M\rVert_*+\langle U,B_1M-R\rangle + \langle V,B_2M-S\rangle + \boldsymbol{w}^\top B_3M \mathbf{1} -\langle G,M \rangle,
\end{equation*}
with $ U, \boldsymbol{w},V\geq 0, G\geq 0$ being dual variables. Further we derive its dual problem:
\begin{equation*}
\begin{split}
    \min_{U,V,\boldsymbol{w},G} \quad & \langle U,R\rangle+\langle V,S \rangle\\
    \text{s.t.} \ \  \quad & \lVert -B_1^\top U - B_2^\top V+B_3^\top \boldsymbol{w} \mathbf{1}^\top+G \rVert_2 \leq 1 \\
    &V \geq 0, \  G\geq 0.
\end{split}  
\end{equation*}
Let the indicator function of a set $C$ be defined as 
\begin{equation*}
    \delta_C(x) = \left\{ \begin{array}{lc}
         0, \quad &x\in C,\\
        +\infty, \quad &x \notin C.
    \end{array}
    \right.
\end{equation*}
Particularly, if $C$ is the set of matrices whose entries are greater than 0, we denote it as $\delta_+(x)$.
By introducing a new variable $H$, we can modify the dual problem to
\begin{equation}\label{eq:modified dual model}
\begin{split}
    \min_{U,V,\boldsymbol{w},G,H} \quad & \langle U,R\rangle+\langle V,S \rangle+\delta_\mathcal{B}(H) +\delta_+(V) +\delta_+(G)\\
    \text{s.t.} \ \  \quad &  -B_1^\top U - B_2^\top V+B_3^\top \boldsymbol{w} \mathbf{1}^\top+G -H=0,
\end{split}  
\end{equation}
where $\mathcal{B}$ refers to the set composed of matrice whose $2$-norm is less than or equal to $1$. For simplicity, we define
\begin{equation*}
    \Gamma(U,V,\boldsymbol{w},G,H)= -B_1^\top U - B_2^\top V+B_3^\top \boldsymbol{w} \mathbf{1}^\top+G -H.
\end{equation*}
Then the augmented Lagrangian function of model \eqref{eq:modified dual model} is given by
\begin{equation*}
    L_\sigma(U,V,\boldsymbol{w},G.H;M)= \langle U,R\rangle+\langle V,S \rangle+\delta_\mathcal{B}(H) +\delta_+(V) +\delta_+(G) + \langle  \Gamma(U,V,\boldsymbol{w},G,H),M\rangle+\frac{\sigma}{2}\lVert\Gamma(U,V,\boldsymbol{w},G,H)\rVert_F^2.
\end{equation*}
We solve the minimizer of the above equation iteratively, in each iteration of which, we minimize it with respect to $U, V, G, U, \boldsymbol{w}, H, \boldsymbol{w}$ in turn, and then take dual ascent for $M$. The $k$-th iteration formula is as follows
\begin{equation}\label{eq:sGs_admm}
	\left\{
	\begin{aligned}
		&U^{(k+\frac{1}{2})}:=\mathop{\arg\min}\limits_{U}\quad \langle U,R\rangle-\langle B_1^\top U,M^{(k)}\rangle+\frac{\sigma}{2}\lVert\Gamma(U,V^{(k)},\boldsymbol{w}^{(k)},G^{(k)},H^{(k)})\rVert_F^2,\\
		&V^{(k+1)}:=\mathop{\arg\min}\limits_{V}\quad \langle V,S \rangle -\langle B_2^\top V,M^{(k)} \rangle+\delta_+(V)+\frac{\sigma}{2}\lVert\Gamma(U^{(k+\frac{1}{2})},V,\boldsymbol{w}^{(k)},G^{(k)},H^{(k)})\rVert_F^2,\\
		&G^{(k+1)}:=\mathop{\arg\min}\limits_{G}\quad \langle G, M^{(k)} \rangle+\delta_+(G) + +\frac{\sigma}{2}\lVert\Gamma(U^{(k+\frac{1}{2})},V^{(k)},\boldsymbol{w}^{(k)},G,H^{(k)})\rVert_F^2,\\
		&U^{(k+1)}:=\mathop{\arg\min}\limits_{U}\quad \langle U,R\rangle+\langle U,M^{(k)}\rangle+\frac{\sigma}{2}\lVert\Gamma(U,V^{(k+1)},\boldsymbol{w}^{(k)},G^{(k+1)},H^{(k)})\rVert_F^2,\\
		&\boldsymbol{w}^{(k+\frac{1}{2})}:=\mathop{\arg\min}\limits_{\boldsymbol{w}}\quad \langle  B_3^\top \boldsymbol{w} \mathbf{1}^\top,M^{(k)}\rangle+\frac{\sigma}{2}\lVert\Gamma(U^{k+1},V^{(k+1)},\boldsymbol{w},G^{(k+1)},H^{(k)})\rVert_F^2,\\
    &H^{(k+1)}:=\mathop{\arg\min}\limits_{H}\quad -\langle H,M^{(k)}\rangle+\delta_\mathcal{B}(H)+ \frac{\sigma}{2}\lVert\Gamma(U^{k+1},V^{(k+1)},\boldsymbol{w}^{(k+\frac{1}{2})},G^{(k+1)},H)\rVert_F^2,\\
    &\boldsymbol{w}^{(k+1)}:=\mathop{\arg\min}\limits_{\boldsymbol{w}}\quad \langle  B_3^\top \boldsymbol{w} \mathbf{1}^\top,M^{(k)}\rangle+\frac{\sigma}{2}\lVert\Gamma(U^{k+1},V^{(k+1)},\boldsymbol{w},G^{(k+1)},H^{(k+1)})\rVert_F^2,\\
		&M^{(k+1)}:=M^{(k)}+\gamma\sigma\Gamma(U^{k+1},V^{(k+1)},\boldsymbol{w}^{(k+1)},G^{(k+1)},H^{(k+1)}).
	\end{aligned}
	\right.
\end{equation}
Ultimately, $M$ at the end of iterations is the solution that we want.
\section{Numerical Experiments}
In this section, we show the performance of SketchDecomp, on a real test bed where stream data with thousands of flows is continually sent from a host of data center $DC_1$ to another one of data center $DC_2$ in ten seconds. The hosts have been synchronized by NTP \cite{mills1991internet}. Packet loss at various scales occurs during transmission. The ground truth of packet loss is known and used to verify our results.
\begin{figure}[h]
     \centering
     \begin{subfigure}{0.4\textwidth}
         \centering
         \includegraphics[width=\textwidth]{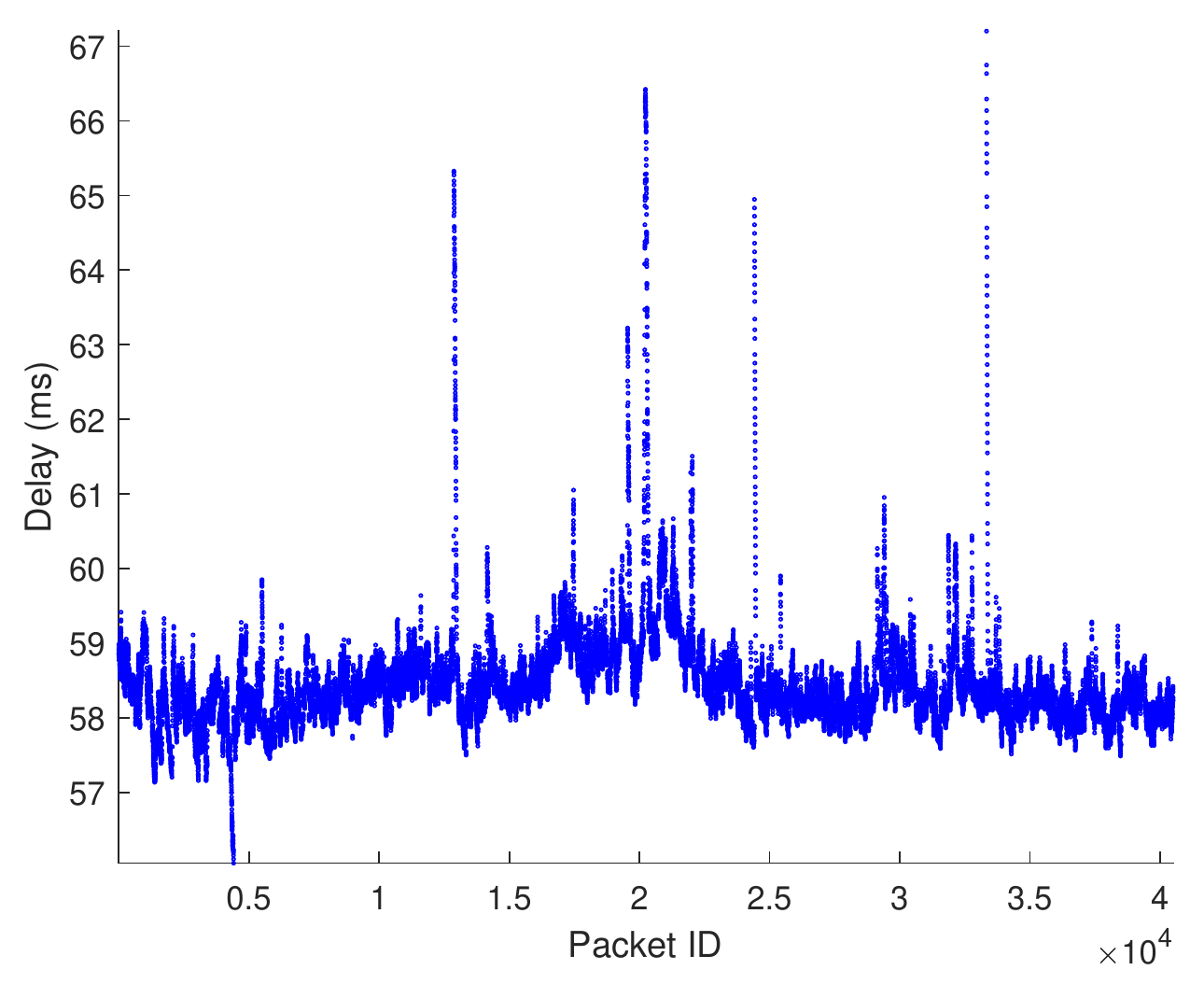}
     \end{subfigure}
     \begin{subfigure}{0.4\textwidth}
         \centering
         \includegraphics[width=\textwidth]{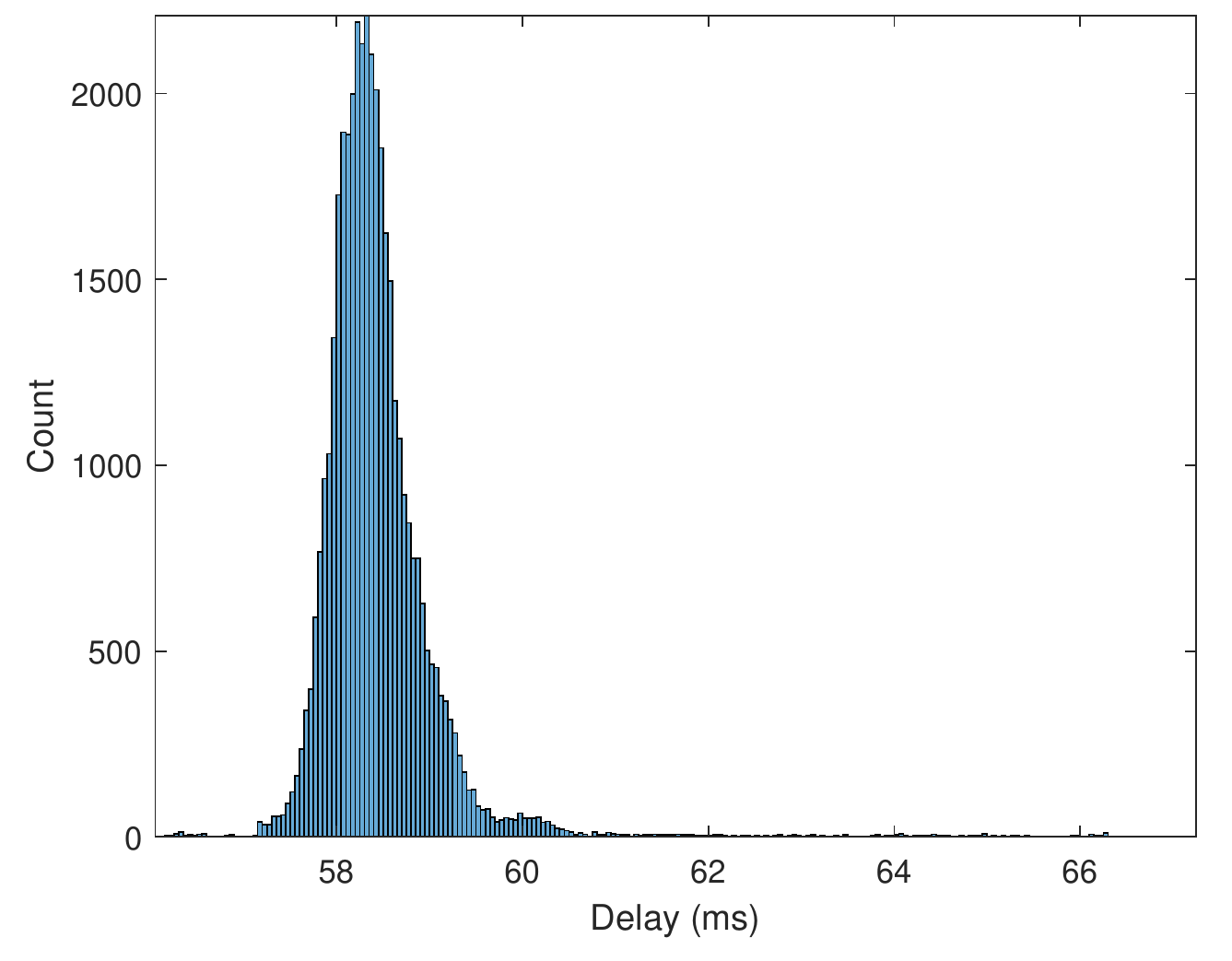}
     \end{subfigure}
     \caption{Actual delays of all packets. \textbf{Left}: delays experienced by different packets. \textbf{Right}: delay distribution (count).}
     \label{fig:delay}
\end{figure}
\subsection{Settings}
We divide the upstream and downstream time into $1000$ time windows of length 10 milliseconds, i.e., $n=1000, \ T=10ms$. CM sketches of depth $4$ ($d=4$) and width $32$ ($w=32$) are adopt to count in each window. Considering the delay variation range is larger than $10ms$ (see Figure \ref{fig:delay}), we set $m=3$. For the final low- rank optimization model solution, we set the parameters of the sGS-ADMM algorithm $\sigma = 1,\  \gamma= 1.618$.

\subsection{Performance}
\begin{figure}[h]
     \centering
     \includegraphics [width=0.8\linewidth]{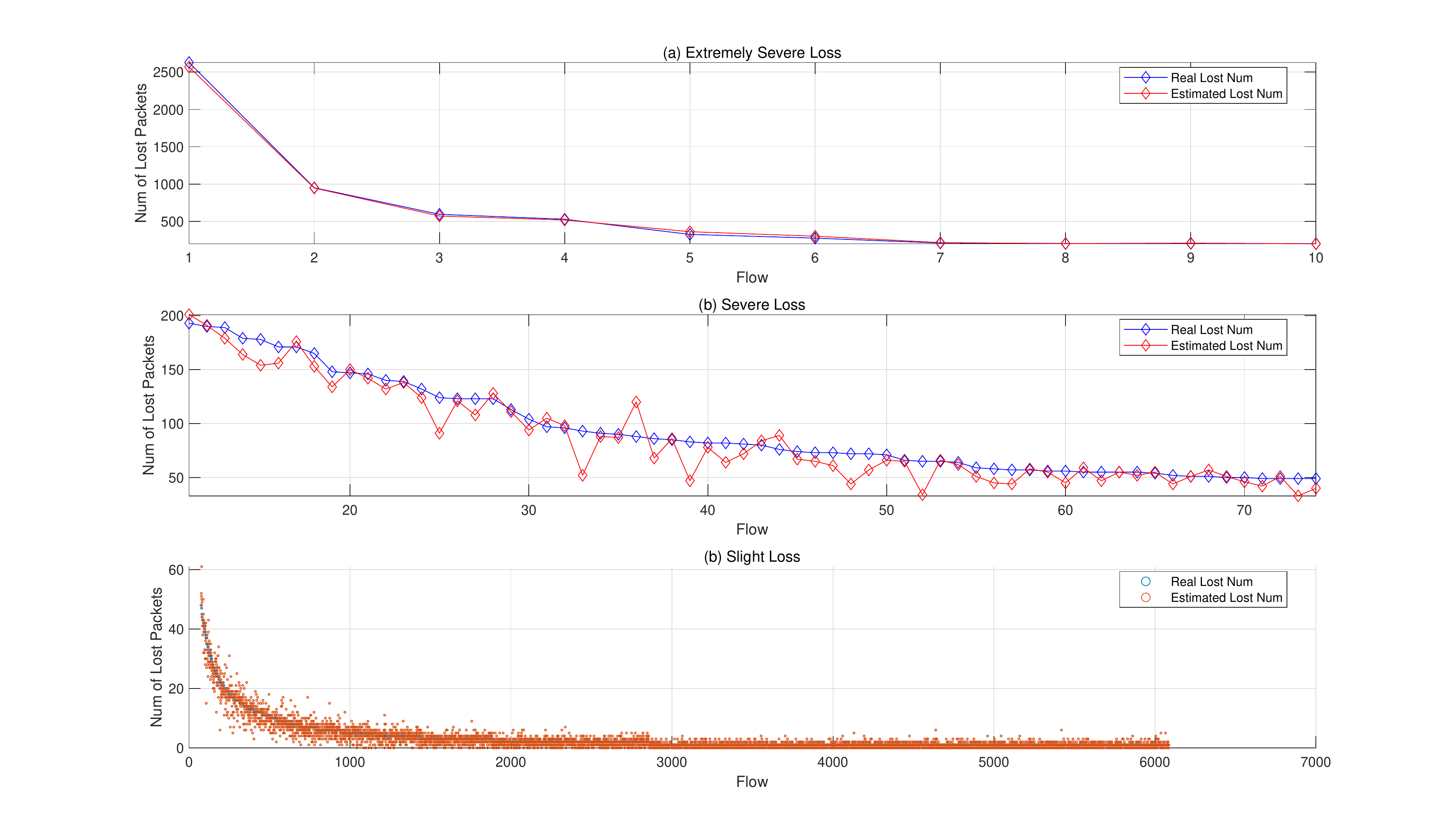}
     \caption{Comparsion of actual and estimated packet loss. \textbf{Top}: Comparison for the flows with extremely severe loss. \textbf{Middle}: Comparison for the flows with severe loss. \textbf{Bottom}: Comparison for the flows with extremely severe loss.}
     \label{fig:loss}
 \end{figure}

\ According to the actual packet loss, we divide flows into three groups : (1) extremely severe loss, (2) severe loss, and (3) slight loss. Tabel \ref{tab:loss} and Figure \ref{fig:loss} compare actual and estimated packet loss numbers in different groups. It is seen that SketchDecomp works significantly well in identifying loss for the flows with extremely severe loss. These flows are exactly of the most concern because they are closely related to network errors. Based on them, data center administrators can perform network diagnostics effectively and address the vulnerability. SketchDecomp also does a satisfying job in severe loss and slight loss, with the ratio being $0.070$ and $0.073$.
\begin{table}[h]
    \centering
    \begin{tabular}{cccc}
        \toprule
         Group & Extremely severe loss & Severe loss & Slight loss\\
         \hline
         Average number of actual loss &621.7 & 93.3 &3.39\\
         Average number of estimated loss &609.7 & 86.8 &3.14\\
         Average error& 12.0 & 6.50 & 0.25\\
         Ratio & 0.019 &  0.070 & 0.073\\
         \hline
    \end{tabular}
    \caption{Comparsion of actual and estimated packet loss. The fourth row is obtained by dividing the third row by the second row.}
    \label{tab:loss}
\end{table}

\section{Conclusion}
We propose a novel packet loss detection method that combines sketch decomposition and matrix optimization. Free from limitations in hardware and inaccessibility of the link, it does a great loss identification job, especially for the flows with much loss. Our experiments confirm it strongly. It is worth emphasizing that previous tools have failed to work on this problem, and our approach fills the gap. This is also the first time anyone has analyzed packet loss from a mathematical point of view.



\end{document}